\documentstyle[epsf,amssymb,twocolumn]{jpsj}

\begin{document}
\sloppy

\title{Fermi Surface and Band Dispersion in La$_{2-x}$Sr$_x$CuO$_4$} 
\author{Akihiro {\sc Ino}$^1$\footnote{E-mail: 
ino@wyvern.phys.s.u-tokyo.ac.jp}, Changyoung {\sc Kim}$^2$, Takashi {\sc 
Mizokawa}$^1$, Zhi-Xun {\sc Shen}$^2$, Atsushi {\sc Fujimori}$^1$ 
Masamitsu {\sc Takaba}$^3$, Kenji {\sc Tamasaku}$^3$\footnote{Present 
address: The Institute of Physical and Chemical Research (RIKEN), 
SPring-8, Kamigori-cho, Hyogo 678-12, Japan.}, Hiroshi {\sc 
Eisaki}$^3$ and Shinichi {\sc Uchida}$^3$}

\inst{$^1$ Department of Physics, University of Tokyo, Bunkyo-ku, 
Tokyo 113-0033, Japan\\
$^2$ Department of Applied Physics and Stanford Synchrotron Radiation 
Laboratory,\\ Stanford University, Stanford, CA 94305, USA\\
$^3$ Department of Superconductivity, University of Tokyo, 
Bunkyo-ku, Tokyo 113-0033, Japan} 

\recdate{February 5, 1999}

\abst{Using angle-resolved photoemission spectroscopy (ARPES), we 
observe the band structure, the Fermi surface and their doping 
dependences in La$_{2-x}$Sr$_x$CuO$_4$.  The results reveal that the 
Fermi surface undergoes a dramatic change: it is holelike and centered 
at $(\pi$,$\pi$) in underdoped ($x=0.1$) and optimally doped 
($x=0.15$) samples as in other cuprates, while it is electronlike and 
centered at (0,0) in heavily overdoped ($x=0.3$) ones.  The peak in 
the ARPES spectra near ($\pi/2$,$\pi/2$) is broad and weak unlike that 
in other cuprates.  In the underdoped and optimally doped samples, a 
superconducting gap ($\Delta = 10 - 15$ meV) is observed near ($\pi$,0).}

\kword{La$_{2-x}$Sr$_x$CuO$_4$, ARPES, Fermi surface, band dispersion, 
superconducting gap, doping dependence}

\def\runtitle{{\sc Letters}}
\def\runauthor{{\sc Letters}}

\maketitle

% Introduction
Among the family of high-$T_{\rm c}$ cuprate superconductors, 
La$_{2-x}$Sr$_x$CuO$_4$ (LSCO) provides unique opportunities to study 
the systematic evolution of the electronic structure with hole doping.  
First, LSCO has a simple crystal structure with single CuO$_2$ layers.  
It has neither Cu-O chains as in YBa$_2$Cu$_3$O$_{7-\delta}$ (YBCO) nor 
complicated structural modulation of the block layers as in 
Bi$_2$Sr$_2$CaCu$_2$O$_{8+\delta}$ (Bi2212).  Second, the hole 
concentration in the CuO$_2$ plane can be controlled over a wide range 
and uniquely determined by the Sr concentration $x$ (and the small 
oxygen non-stoichiometry).  One can therefore investigate the doping 
dependence of the electronic structure continuously from the heavily 
overdoped limit ($x\sim0.35$) to the undoped insulator ($x=0$) in the 
same system.  This information would be highly useful to critically 
check existing theories of electron correlations and superconductivity 
in the CuO$_2$ plane.  Indeed, the doping dependences of thermodynamic 
and transport properties have been extensively studied for the LSCO 
system~\cite{Takagi,Nakano,Hwang,LSCOspecheat1,LSCOspecheat2}.

Angle-resolved photoemission spectroscopy (ARPES) is a powerful method 
to probe the electronic structure of low-dimensional systems.  In 
particular, band structures, Fermi 
surfaces~\cite{Olson,Bi2212a,Bi2212b,Bi2201-FS,YBCO-FS,AndNCCO,NCCO} 
and superconducting and normal-state gaps in the high-$T_{\rm c}$ 
cuprates~\cite{ShenGap,Marshall,Loeser,Ding,Harris,Bi2201,YBCO} have 
been observed by ARPES. However, most of the ARPES experiments have 
focused on the Bi2212 system and its family compounds; ARPES studies 
of LSCO have been hindered probably because LSCO is difficult to cleave and 
its surface is not as stable as that of Bi2212 under an ultrahigh 
vacuum.

We have recently focused on the LSCO system and carried out a series 
of photoemission studies~\cite{Chem,AIPES}.  The angle-integrated 
photoemission (AIPES) spectra of LSCO  show a broad feature (at 
$\sim -100$ meV) and a suppression of the density of states at the 
Fermi level ($E_{\rm F}$) in the underdoped region~\cite{AIPES}.  In the 
present study, we have overcome the experimental difficulties in the 
ARPES of LSCO using high-quality single crystals, and have measured 
ARPES spectra of underdoped ($x=0.1$), optimally doped ($x=0.15$) and 
heavily overdoped ($x=0.3$) samples in order to investigate the doping 
dependence of the electronic structure of the CuO$_2$ plane, in 
particular, the shape of the Fermi surface and the band dispersions.

%  Experimental
Single crystals of LSCO ($x=0.1, 0.15$ and 0.3) were grown by the 
traveling-solvent floating-zone method.   ARPES measurements were 
carried out at beamline 5-3 of Stanford Synchrotron Radiation 
Laboratory (SSRL), using incident photons with energies of 22.4 eV and 
29 eV. The wave vector and the electric vector of the incident photons 
and the sample surface normal were kept in the same plane and the angle 
of incidence was 45$^\circ$.  The total energy resolution was 
approximately 45 meV and the angular resolution was $\pm 1$ degree.  
The spectrometer was kept in an ultrahigh vacuum better than $5\times 
10^{-11}$ Torr during the measurements and the samples were cleaved 
{\it in situ}.  Since the surface degraded rapidly at high 
temperatures, the measurements were done only at low temperatures ($T 
\sim 15$ K).  The cleanliness of the surface was confirmed by the 
absence of a hump at energy $\sim -9.5$ eV and a shoulder in the 
valence band at $\sim -5$ eV. All the spectra presented here were 
taken within 10 hours and mostly five hours after cleaving.  The position 
of the Fermi level was calibrated with gold spectra for every 
measurement and the experimental uncertainty of the energy calibration 
was about $\pm 2$ meV.

\begin{figure}[t]
 \epsfxsize=85mm \epsfbox{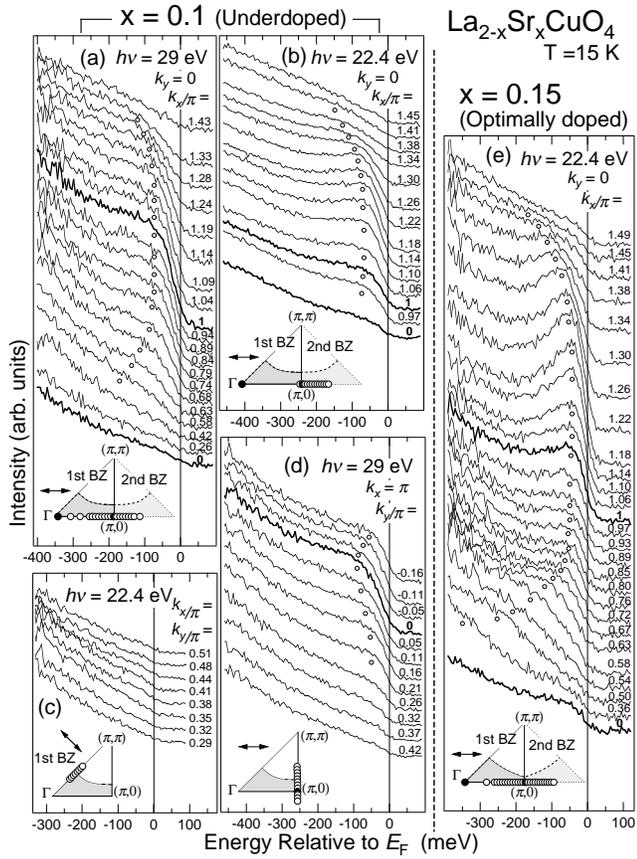} \vspace{-1pc}
 \caption{ARPES spectra of underdoped ($x=0.1$) and optimally doped 
 ($x=0.15$) La$_{2-x}$Sr$_x$CuO$_4$ near the Fermi level.  Insets show 
 measured points in the momentum space and the polarization of 
 incident photons (arrows).}
\label{x=0.1}
\end{figure}

%spectra and FS crossing points
ARPES spectra for underdoped ($x=0.1$) and optimally doped ($x=0.15$) 
LSCO are shown in Fig.~\ref{x=0.1}.  Although the dispersive features 
are broad, an angular dependence is identified.  As one goes from 
(0,0) to ($\pi$,0) in the first Brillouin zone (BZ) or from ($2\pi$,0) 
to ($\pi$,0) in the second BZ [Figs.~\ref{x=0.1}(a) and 
\ref{x=0.1}(b)], a broad ``quasiparticle'' (QP) peak emerges from 
lower energies and then stays somewhat below $E_{\rm F}$ around the 
($\pi$,0) point, never crossing the Fermi surface.  The midpoint of 
the leading edge is below $E_{\rm F}$ ($\sim -20$ meV) at 
$\sim$($\pi$,0).  As for Fig.~\ref{x=0.1}(b), the peak intensity 
decreases in going from (1.2$\pi$,0) to ($\pi$,0).  This is probably 
due to a matrix-element effect peculiar to the incident photon energy 
of $h\nu = 22.4$ eV because such an intensity modulation is not seen 
in the spectra taken at another photon energy $h\nu = 29$ eV, as shown 
in Fig.~\ref{x=0.1}(a).  The dispersion of the QP peak is very weak 
around the ($\pi$,0) point, meaning that there is a ``flat band'', 
namely, an extended saddle point around ($\pi$,0), similar to that in 
other cuprates~\cite{Bi2201-FS}.  On the other hand, as one goes from 
($\pi$,0) towards ($\pi$,$\pi$) [Fig.~\ref{x=0.1}(d)], the broad peak 
and its leading edge further approach $E_{\rm F}$ and then the peak 
disappears around $\sim$($\pi$,0.2$\pi$).  Since a superconducting gap 
is opened on the Fermi surface at this temperature ($T\sim 15$ 
K$<T_{\rm c}$), the leading-edge midpoint stays below $E_{\rm F}$ 
($\sim -8$ meV) even for the peak closest to $E_{\rm F}$.  It has been 
indeed reported that the minimum-gap locus in the superconducting 
state coincides with the Fermi surface in the normal 
state~\cite{Campuzano}.  Therefore, we may conclude that the band 
crosses an underlying Fermi surface around $\sim$($\pi$,0.2$\pi$).  
(If the energy gap of underdoped LSCO remain opened in the normal 
state as indicated for the underdoped 
Bi2212~\cite{Loeser,Ding,Harris}, no real Fermi surface exists in the 
underdoped LSCO.) As for the (0,0)$\to$($\pi$,$\pi$) cut 
[Fig.~\ref{x=0.1}(c)], no QP peak is identified in the present 
spectra.  However, the fact that the QP band is below $E_{\rm F}$ at 
($\pi$,0) implies that the Fermi surface underlying the 
superconducting gap is holelike and centered at ($\pi$,$\pi$) for 
$x=0.1$, as in other cuprates studied previously such as Bi2212, 
Bi2201, YBCO and 
Nd$_{2-x}$Ce$_x$CuO$_4$~\cite{Bi2212a,Bi2212b,Bi2201-FS,YBCO-FS,AndNCCO,NCCO}.

Spectra for $x=0.15$ show a behavior similar to those for $x=0.1$: the 
QP peak and the leading-edge midpoint stay below $E_{\rm F}$ in 
(0,0)$\to$($\pi$,0) as shown in Fig.~\ref{x=0.1}(e) and the peak 
disappears in going from ($\pi$,0) to ($\pi$,$\pi$), indicating a 
holelike underlying Fermi surface, as in $x=0.1$.  The leading-edge 
midpoint reaches about -3 meV at the closest to $E_{\rm F}$.  The peak 
intensity is pronounced around (1.2$\pi$,0) because of the 
matrix-element effect peculiar to $h\nu = 22.4$ eV, as in $x=0.1$ 
[Fig.~\ref{x=0.1}(b)] and some other cuprates~\cite{YBCO-FS}.  
Note, however, that no decrease in the peak intensity is found in 
$\sim$(0.8$\pi$,0)$\to$($\pi$,0) in the first BZ for $x=0.15$.

\begin{figure}[t]
 \epsfxsize=85mm \epsfbox{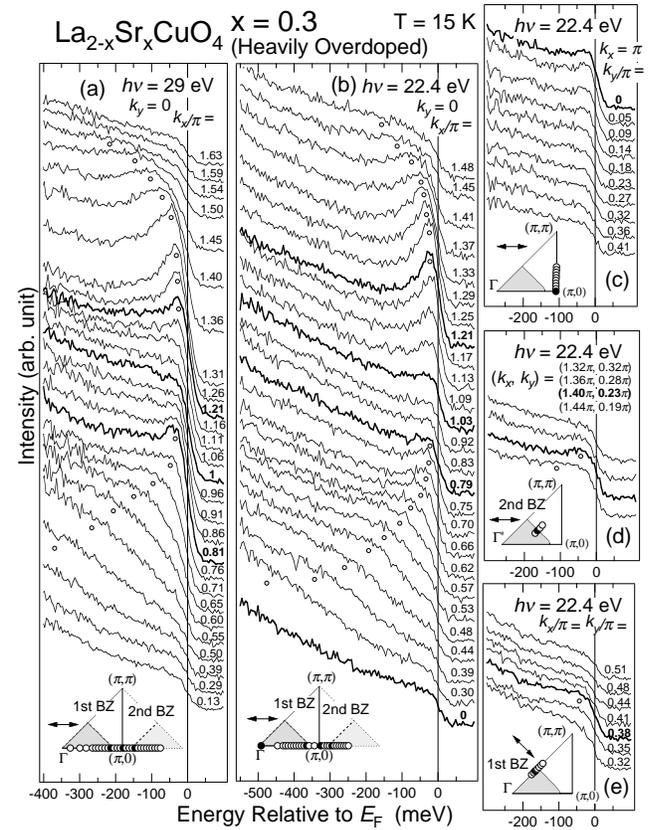}  \vspace{-1pc}
 \caption{ARPES spectra of heavily overdoped ($x=0.3$) 
 La$_{2-x}$Sr$_x$CuO$_4$ near the Fermi level.  Insets show 
 measured points in the momentum space and the polarization of 
 incident photons (arrows).}
\label{x=0.3}
\end{figure}

ARPES spectra for heavily overdoped ($x=0.3$) LSCO are shown in 
Fig.~\ref{x=0.3}.  In going from (0,0) to ($\pi$,0) or from ($2\pi$,0) 
to ($\pi$,0) [Figs.~\ref{x=0.3}(a) and \ref{x=0.3}(b)], the QP peak 
moves upwards, reaches $E_{\rm F}$ around $\sim$($0.8\pi$,0) or 
$\sim$($1.2\pi$,0), then decreases in its intensity and almost 
disappears at ($\pi$,0).  Here, the leading-edge midpoint also reaches 
$\sim +6$ meV above $E_{\rm F}$ at $\sim$(1.2$\pi$,0), suggesting that 
the QP band has indeed reached the Fermi level.  That should be 
contrasted with the spectra observed for Bi2212, where the 
leading-edge midpoint is always below $E_{\rm F}$ in the 
(0,0)$\to$($\pi$,0) cut~\cite{Bi2212a,Loeser}.  The intensity decrease 
at $\sim$($0.8\pi$,0) occurs in both the first and second BZs and 
in the spectra taken at both $h\nu = 29$ eV and $h\nu = 22.4$ eV. 
Hence, it is difficult to explain all the observed intensity decreases 
unless a Fermi-surface crossing occurs there besides the effect of 
matrix elements, because the matrix-element effect is dependent on the 
energy of incident photons and is different between the first and 
second BZs.  As for the spectra taken at $h\nu = 29$ eV, the intensity 
decrease at $\sim$(0.8$\pi$,0) and $\sim$(1.2$\pi$,0) observed for 
$x=0.3$ [Fig.~\ref{x=0.3}(a)] is not found for $x=0.1$ 
[Fig.~\ref{x=0.1}(a)], distinguishing the two compositions.  
Furthermore, in the spectra taken at $h\nu = 22.4$ eV in the first BZ, 
the QP peak seen at (0.75$\pi$,0) almost disappears at ($0.92\pi$,0) 
for $x=0.3$ [Fig.~\ref{x=0.3}(b)], in contrast to the spectra for 
$x=0.15$ [Fig.~\ref{x=0.1}(e)]. Therefore, we have concluded that 
the QP band crosses the Fermi surface around (0.8$\pi$,0) for $x=0.3$.  
Close inspection of the spectra [Figs.~\ref{x=0.3}(b) and \ref{x=0.3}(c)] 
reveals that a small part of the QP peak weight appears to remain 
below $E_{\rm F}$ at $\sim$($\pi$,0), suggesting that the ``flat 
band'' stays only slightly above $E_{\rm F}$ at ($\pi$,0).  In the 
presence of strong electron correlation, the QP peak is no more a 
single peak and part of the spectral weight is distributed on the 
other side of $E_{\rm F}$.  This also explains the intensity drop in 
going from ($\pi$,0) to ($\pi$,$\pi$) as seen in Fig.~\ref{x=0.3}(c).  
Along the (0,0)$\to$($\pi$,$\pi$) cut [Fig.~\ref{x=0.3}(d)], although 
the peak is not clearly identified, the edge intensity slightly 
increases at $\sim$($0.38\pi$,$0.38\pi$) and then drops at 
$\sim$($0.44\pi$,$0.44\pi$), a behavior indicative of  Fermi-surface 
crossing.  This intensity variation was reproduced in several 
measurements.  The QP peak near ($\pi/2$,$\pi/2$) is broad and weak 
compared to the peak near ($\pi$,0), in contrast to other cuprates 
showing clear QP peaks around 
($\pi/2$,$\pi/2$)~\cite{Bi2212a,Bi2212b,Marshall,Loeser,Shen&Schrieffer}. 
 To summarize, the observations indicate that the Fermi surface is 
electronlike and centered at the (0,0) point for $x=0.3$, unlike the 
other cuprates studied so far.

\begin{figure}[t]
 \epsfxsize=60mm \centerline{\epsfbox{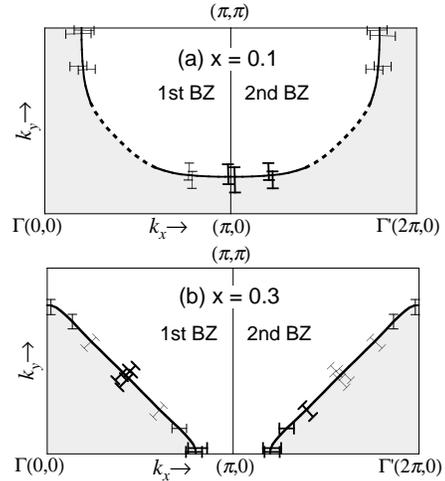}}  \vspace{-1pc} 
 \caption{Fermi surfaces of La$_{2-x}$Sr$_x$CuO$_4$ for $x=0.1$ (a) 
 and $x=0.3$ (b).  Observed Fermi-surface crossings are denoted by 
 thick error bars.  Thin error bars indicate Fermi-surface 
 crossings folded by symmetry.  As for $x=0.1$, minimum-gap locus 
 are taken as Fermi-surface crossings and the dotted curve is 
 tentatively drawn so that the area enclosed by the Fermi surface 
 is $\sim90\%$ of the half BZ area.}
\label{k-space}
\end{figure}

Thus the Fermi surface undergoes a dramatic change from holelike to 
electronlike between $x=0.15$ and $x=0.3$, as the flat band around 
($\pi$,0) moves from below $E_{\rm F}$ to above $E_{\rm F}$.  The 
change in the Fermi-surface topology may be related to the observation 
that the sign of the Hall coefficient changes from positive to 
negative around $x=0.25$ in LSCO~\cite{Tamasaku,Hall} and has the same 
tendency as that expected from the local-density-approximation 
(LDA) band-structure calculation of La$_2$CuO$_4$ by shifting the 
Fermi level as in the rigid band model~\cite{Mattheiss,bandcalcLSCO}.  
The behavior of the ($\pi$,0) flat band upon doping is consistent with 
numerical studies of the Hubbard model~\cite{Duffy,Preuss}.  
Figure~\ref{k-space} shows the Fermi surfaces suggested from the 
present experiments for $x=0.1$ and 0.3.  The area enclosed by the 
Fermi surface is $71\pm3\%$ of the half BZ area for $x=0.3$, which 
agrees well with the number of electrons $1-x$ as expected from the 
Luttinger sum rule, indicating a ``large Fermi surface''.  As for 
$x=0.1$, the Fermi surface near ($\pi/2$,$\pi/2$) (dashed curve) has 
been tentatively drawn in Fig.~\ref{k-space} so that the enclosed area 
is consistent with the Luttinger sum rule.  According to 
Fig.~\ref{k-space}, the Fermi surface for $x=0.3$ seems to be almost 
square and have a large straight portion around ($\pi/2$,$\pi/2$).  
Recently, Fermi-surface nesting~\cite{nesting} and short-range stripe 
order~\cite{Tranquada} have been proposed as the origins of the 
incommensurate peaks in the dynamical magnetic structure factor 
$S(\textbf{\textit{q}},\omega)$ observed by inelastic neutron 
scattering.  However, the Fermi-surface nesting would not explain why 
the the incommensurate peaks are smeared out in the overdoped region 
($x>0.25$), as recently observed~\cite{INS0.3} in the presence of the 
straight Fermi surface for $x=0.3$.

%% speculations

%broad features

The width of the QP peak around ($\pi$,0) for LSCO is almost the same 
as that for Bi2212 with corresponding hole 
doping~\cite{Bi2212a,Bi2212b,Marshall,Loeser,Shen&Schrieffer}, even 
though the dispersive features are weaker for LSCO than for Bi2212.  
The QP peak becomes broader as $x$ decreases from the overdoped to the 
underdoped regions.  The broad peak on the flat band around ($\pi$,0) 
may be related to the broad feature in the AIPES spectra, where the 
spectral intensity starts to decrease towards $E_{\rm F}$ in the 
underdoped LSCO~\cite{AIPES}, that is, what is called a high-energy 
pseudogap or a weak pseudogap~\cite{Pines} on an energy scale ($\sim 
100$ meV) much larger than that of the superconducting gap and the 
``normal-state gap'' ($\lesssim 20$ 
meV)~\cite{Loeser,Ding,Harris,Bi2201,YBCO}.  Both the broad ($\pi$,0) 
peak in the ARPES spectra and the high-energy pseudogap in the AIPES 
spectra have nearly the same energy scales, that is, the order of the 
superexchange energy $J\sim100$ meV, and their energies also increase 
as $x$ decreases, following the development of antiferromagnetic 
correlations~\cite{AIPES}.  The QP peak line shape and its doping 
dependence in the (0,0)$\rightarrow$($\pi$,0) cut of LSCO are also 
consistent with the trend seen among Bi2212, Sr$_2$CuO$_2$Cl$_2$ and the 
$t-t'-t''-J$ model calculation~\cite{Shen&Schrieffer,Kim}.  These 
observations suggest that the broad line shape of the ARPES spectra 
around ($\pi$,0) originates from antiferromagnetic correlations.

\begin{figure}[t]
 \epsfxsize=61mm \centerline{\epsfbox{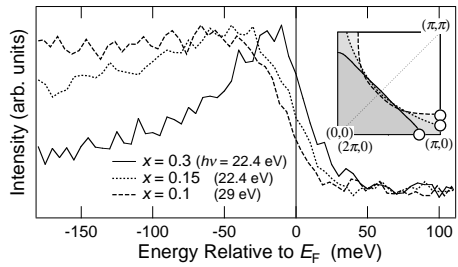}}  \vspace{-1pc}
 \caption{ARPES spectra for momenta on the Fermi surface (minimum-gap 
 locus) near ($\pi$,0) shown by open circles in the inset.  
 For each composition $x$, the spectrum at (0,0) has been subtracted as 
 the angle-independent background.}
\label{scgap}
\end{figure}

%energygap
The magnitude of the superconducting gap may be roughly estimated from 
the shift of the leading edge on the Fermi surface 
(Fig.~\ref{scgap})~\cite{Harris,Bi2201}.  For the non-superconducting 
sample ($x=0.3$), the leading-edge midpoint is pushed above $E_{\rm F}$ 
($\sim +6$ meV) at the Fermi-surface crossing $\sim$($1.2\pi$,0) due 
to the finite instrumental resolution ($\sim 45$ meV).  On the other 
hand, the leading-edge midpoint is below $E_{\rm F}$, $\sim -8$ and $\sim 
-3$ meV for $x=0.1$ and 0.15, respectively, on the Fermi surface near 
($\pi$,0).  Accordingly, the relative shift of the leading edge is 
about $10 - 15$ meV on the Fermi surface near ($\pi$,0) for $x=0.1$ 
and 0.15, and is smaller than the typical value $\sim 25$ meV for 
Bi2212 and YBCO~\cite{Loeser,Harris,YBCO}, but is similar to that for 
Bi$_2$Sr$_2$CuO$_{6+\delta}$ (Bi2201)~\cite{Bi2201}, which also has 
single CuO$_2$ planes.  These leading-edge shifts scale well with the 
maximum $T_{\rm c}$ of the systems.  The superconducting gap of $\Delta= 10 
- 15$ meV estimated here is roughly consistent with Raman 
scattering~\cite{Raman}, tunneling~\cite{Tunnel} and inelastic
neutrons scattering~\cite{INS} results for LSCO.

%summary
In conclusion, we have successfully observed band dispersions and 
Fermi surfaces in overdoped and underdoped LSCO. We found that the 
holelike underlying Fermi surface centered at ($\pi$,$\pi$) for 
$x=0.1$ and 0.15 is converted into the electronlike Fermi surface 
centered at (0,0) for $x=0.3$ (heavily overdoped).  The peak in the 
ARPES spectra near ($\pi/2$,$\pi/2$) is quite broad and weak compared 
to that of other cuprates.  In the underdoped and optimally doped 
samples, a superconducting gap ($\Delta=10-15$ meV) is observed on the 
Fermi surface at $\sim$($\pi$,0.2$\pi$).  More studies are necessary 
to identify the anisotropy, doping dependence and temperature 
dependence of the superconducting gap.

%%acknoledgement
This work was supported by a Grant-in-Aid for Scientific Research from 
the Ministry of Education, Science, Sports and Culture of Japan, the 
New Energy and Industrial Technology Development Organization (NEDO), 
Special Coordination Fund for Promoting Science and Technology from the
Science and Technology Agency of Japan, the U.~S.~DOE, Office of Basic 
Energy Science and Division of Material Science.  Stanford Synchrotron 
Radiation Laboratory is operated by the U.~S.~DOE, Office of Basic 
Energy Sciences, Division of Chemical Sciences.

\end{document}